\documentclass[12pt]{iopart}

\usepackage{iopams}
\usepackage{amssymb}
\usepackage{amsfonts}
\usepackage{bm}
\usepackage{epsfig}
\usepackage{epsf}
\usepackage[]{graphicx}

\usepackage{color}

\usepackage{cite}
\usepackage{xspace}

\newcommand {\bfr} {{\bf r}}
\newcommand {\bfv} {{\bf v}}

\newcommand {\bfE} {{\bf E}}

\renewcommand {\d} {{\rm d}}

\newcommand {\E} {\varepsilon}

\newcommand {\bfnabla} {\boldsymbol{ \nabla}}
\newcommand {\aTF}  {a_{\rm TF}}

\newcommand{\MBNExplorer} {\textsc{MBN Explorer}\xspace}

\newcommand{\Lp}{L_{\rm p}}
\newcommand{\Ld}{L_{\rm d}}
\newcommand{\tb}{\theta_{\rm b}}

\begin{document}

\title[Simulations of channeling and radiation processes
in thin Si and Ge crystals]{Atomistic modelling of
electron propagation and
radiation emission in oriented bent ultra-thin Si
and Ge crystals}

\author{V. V. Haurylavets$^1$,
V. K. Ivanov$^2$,
A. V. Korol$^3$, and A. V. Solov'yov$^3$}

\address{$^1$Research Institute for Nuclear Problems,
Belarusian State University, Bobruiskaya street, 11, Minsk 220030, Belarus}

\address{$^2$Peter The Great St.Petersburg Polytechnic University,
Polytekhnicheskaya 29, St Petersburg, 194251, Russia}

\address{$^3$MBN Research Center, Altenh\"{o}ferallee 3,
60438 Frankfurt am Main, Germany}



\begin{abstract}
Computational modelling of passage of high-energy electrons through
crystalline media is carried out by means of the relativistic
molecular dynamics.
The results obtained are compared with the experimental data
for 855 MeV
electron beam incident on oriented bent ultra-thin (15 microns)
silicon and germanium crystals.
The simulations have been performed for the geometries of the
beam--crystal orientation that correspond
(i) to the channeling regime
and (ii) to the volume reflection.
A comparison with the experiment is carried out in terms of
angular distributions of the electrons deflected by the crystals
bent with different curvature radii as well as of the spectra of
the emitted radiation.
For both crystals a good agreement between the simulated and
experimentally measured data is reported.
The origin of remaining minor discrepancies between theory
and experiment is discussed.
\end{abstract}


\section{Introduction \label{Introduction}} %

The passage of a charged particle through a crystalline
environment depends largely on the orientation of the
particle's momentum and a crystallographic direction.
Lindhard has demonstrated \cite{Lindhard} that regularity in the
atomic positions in a crystal can result in a specific channeling
motion when a particle moves along a crystallographic plane
 or axis experiencing correlated interactions with the atoms.
Since then, the passage of beams of ultra-relativistic charged
particles through oriented crystals (channeling phenomenon
included) has become a broad field of research
\cite{BiryukovChesnokovKotovBook,UggerhojRPM}.
The knowledge acquired in theoretical and
experimental investigations has led to a number of applications,
either already implemented or potential ones, the realisation
of which requires further efforts.
Examples of the former include
beam steering \cite{MazzolariEtAl:PRL_v112_135503_2014,
MazzolariEtAl:EPJC_v78_720_2018,
WienandsEtAl:PRL_v114_074801_2015},
collimation \cite{Scandale_EtAl-PRB_v692_p78_2010},
focusing \cite{Scandale_EtAl-NIMB_v446_p15_2019},
and
extraction \cite{BiryukovChesnokovKotovBook}.
Oriented crystals of different geometries (linear, bent,
periodically bent) exposed to the beams of ultra-relativistic
electrons and positrons can potentially serve as novel intensive
gamma-ray crystal-based light sources (CLS) operating in the
MeV-GeV photon energy range
\cite{CLS-book_2022,SushkoEtAl:EPJ_v76_166_2022,%
AVK-AVS:NIMB_v537_p1_2023}.
In addition to the channeling radiation
\cite{Kumakhov:PL_v57A_17_1976}, the particles channeling in
bent crystals can emit synchrotron-like radiation
due to the circular motion along the bent planes
\cite{KaplinVorobiev:PLA_v67_p135_1978,
TaratinVorobiev:NIMB_v31_p551_1988}.
Motion along periodically bent planes gives rise to the intensive
undulator-type radiation \cite{KSG1998,SecondEdition}.
To a great extent, intensity of the radiation and,
consequently, characteristics of CLS (number of photons,
brilliance), depend on the magnitude of the
so-called dechanneling length
$L_{\rm d}$, i.e., the average distance over which a particle
moves in the channeling mode of motion before it leaves the
channel due to uncorrelated collisions with atoms (the
dechanneling process).
In turn, $L_{\rm d}$ is determined by the energy and charge of
the projectile particle, crystal orientation, and type of crystal
atoms, and bending curvature (see, e.g., a review paper
\cite{KorolSushkoSolovyov:EPJD_v75_p107_2021}).
This parameter, which characterizes channeling efficiency, can
be measured experimentally and/or calculated by means of accurate
simulations of particles' passage through oriented crystals.

In recent years, a series of experiments has been carried out at
different accelerator facilities aimed at investigating
channeling and radiation emission phenomena in
linear \cite{BackeLauth:NIMB_v335_p24_2015},
bent
\cite{
MazzolariEtAl:PRL_v112_135503_2014,
WienandsEtAl:PRL_v114_074801_2015,
BandieraEtAl:PRL_v115_025504_2015,
WistisenEtAl:PR-AB_v19_071001_2016,Wienands_EtAl-NIMB_v402_p11_2017,
SytovEtAl:EPJC_v77_901_2017,Bagli_EtAl-EPJC_v77_71_2017,
Scandale_EtAl-EPJC_v79_99_2019,
BandieraEtAl:EPJC_v81_284_2021},
and periodically bent crystals
\cite{BackeEtAl:JPConfSer_v438_012017_2013,
BagliEtAl:EPJC_v74_3114_2014,
WistisenEtAl:PRL_v112_254801_2014,
UggerhojWistisen:NIMB_v355_p35_2015}.

In this paper, we present an independent analysis of the passage
of ultra-relativistic electrons through oriented thin silicon and
germanium crystals and of the emitted radiation.
The analysis is based on the results of simulations performed
within the
framework of relativistic classical molecular dynamics
by means of the \textsc{MBN Explorer} software package
\cite{MBN_Explorer_2012,MBNExplorer_Book,mbn-explorer-software}
 and a supplementary special multitask software toolkit
\textsc{MBN Studio} \cite{MBN_Studio_2019}.
The results obtained are compared to the experimental data on
angular distributions of electrons after passing through silicon
and germanium crystalline targets presented in
Ref. \cite{SytovEtAl:EPJC_v77_901_2017}
and on the measured emission spectra taken from
Ref. \cite{BandieraEtAl:EPJC_v81_284_2021}.
Both experiments were carried out at the MAinzer MIcrotron (MAMI)
facility with a 855 MeV electron beam.
The facility generates highly collimated beam so that in the
simulations its angular divergence has been ignored.
The beam was incident on thin silicon and germanium bent crystals.
The crystals used were of high quality with a very low
concentration of defects.
By means of a specially designed holder
\cite{SalvadorEtAl:JINST_v13_C04006_2018}
the same crystalline sample was bent remotely to achieve different
bending curvatures.
Beyond the target, the electrons were deflected by magnets and
thus separated from the emitted photons.
This allowed  one to measure the angular distributions along with
the emission spectra.
More details on the experimental setup are given in Ref.
\cite{HaurylavetsEtAl:EPJPlus_v137_34_2022}.

In Section \ref{Methodology}, we overview the methodology utilized
to carry out the simulations.
The parameters of the targets as well as the beam-crystal
alignments used in the experiments are described in Section
\ref{Alignment}.
In Section \ref{CaseStudies}, the numerical results obtained are
compared with the experimental data collected at MAMI.
Section \ref{Conclusions} summarizes the conclusions of this work
and presents future perspectives.

\section{Methodology \label{Methodology}}

In this paper the method of relativistic classical
molecular dynamics \cite{MBN_ChannelingPaper_2013},
implemented in the \MBNExplorer package
\cite{MBN_Explorer_2012,MBNExplorer_Book,MBN_Studio_2019,
mbn-explorer-software}, is employed to model the motion of
charged ultra-relativistic particles in an electrostatic field
of the crystalline medium.
This approach implies generation of a large number $N$ of
statistically
independent trajectories of projectile particles that
can be analyzed further to carry out quantitative characterization
of the particles' motion as well as of the radiation emitted.

To model the motion an ultra-relativistic particle
of mass $m$, charge $q$ and energy $\E$
in an atomic environment, the following relativistic equations of
motion are integrated numerically:
\begin{eqnarray}
\dot{\bfr} = \bfv,
\qquad
\dot{\bfv} =
{q \over m \gamma}
\left(\bfE -  {\bfv\left(\bfE \cdot \bfv\right) \over c^2} \right)
\label{Methodology:eq.01}
\end{eqnarray}
where
$\bfr=\bfr(t)$ and $\bfv=\bfv(t)$ are the instantaneous
position vector and velocity of the projectile particle,
$\gamma = \E/mc^2$
denotes the relativistic factor.
The electric field at point $\bfr$
is calculated as $\bfE =-\bfnabla \phi(\bfr)$
with $\phi({\bfr})$ standing for the field's potential.
This quantity is calculated as the sum of
potentials of individual atoms located at points $\bfr_i$:
\begin{eqnarray}
\phi(\bfr) = \sum_i \phi_{\rm at}
\left(\left|\bfr - \bfr_i\right|\right)\,.
\label{Methodology:eq.02}
\end{eqnarray}
The atomic potentials can be computed within the frameworks of the
approximations due to Moli\'{e}re \cite{Moliere}
and Pacios \cite{Pacios}.
The latter is based on the solutions of the Hartree-Fock
equations, therefore, it is more accurate especially at
distances larger than the average atomic radius that can be
estimated as the Thomas-Fermi
radius $a_{\rm TF}$.
At distances less than $a_{\rm TF}$ both approximations
provide the same result in contrast to frequently used
Doyle-Turner scheme \cite{DoyleTurner1968}.
A comparative analysis of the atomic and interplanar potentials
for Si an Ge calculated within the frameworks of these
approximations is presented in Section \ref{Appendix}.

For a neutral atom, the potential $\phi(\bfr)$ decreases
rapidly at the distances $r\gg a_{\rm TF}$.
Therefore, at each step of integration of the equations of motion
(\ref{Methodology:eq.01}), the sum in (\ref{Methodology:eq.02})
can be truncated by accounting only for those
atoms that are located inside the sphere of a specified cut-off
radius $\rho$ with the center at $\bfr$.
Typically, $\rho$ is chosen to be tens times larger than
the average atomic radius.
The search for such atoms is carried out by means of the
linked cell algorithm  implemented in the \MBNExplorer package.
In the course of particle's passage through a crystal, the
crystalline environment around the particle is generated by means
of a dynamic simulation box which moves following the particle.
Inside the box, the nodal positions are generated in accordance
with the crystal lattice and accounting for the transformations
that modify the structure to achieve desired geometry of the
crystal (e.g., linear, bent, periodically bent crystals).
The positions of the atoms are generated accounting for random
displacement from the nodes due to thermal vibrations.
More detailed description on the algorithms used to compute the
trajectories at various scales,
including macroscopically large ones, are presented
in Refs.
\cite{MBN_Explorer_2012,MBN_ChannelingPaper_2013,MBNExplorer_Book,
KorolSushkoSolovyov:EPJD_v75_p107_2021}.

To calculate a trajectory,  the values of the
transverse coordinates and velocities at the crystal entrance
are generated randomly accounting for
the crystal orientation with respect to the incident beam and for
the beam emittance and size.
Allowing also for randomness in positions of the lattice atoms due
to thermal vibrations, one concludes that each trajectory
corresponds to a unique crystalline environment.
Therefore, all simulated trajectories are statistically
independent and
can be analyzed further to quantify the process of interest.
For each trajectory simulated the spectral distribution $\d E_j$
of the radiant energy emitted within a specified cone along the
incident beam can be evaluated numerically following the procedure
described in detail in Ref. \cite{MBN_ChannelingPaper_2013}.
To calculate the total emission spectrum (per particle) one
averages $\d E_j$ over all trajectories:
$\d E=N^{-1}\sum_{j=1}^N \d E_j$, where
$N$ is the number of trajectories.

The simulations have been performed for 855 MeV electrons
passing through bent single silicon and germanium crystals.
We aimed at quantitative description  of orientational effects
(channeling, volume reflection) that reveal themselves in
angular distributions of deflected particles as well as in
the spectra of emitted radiation.
The dependencies of the distributions and the spectra on the
atomic number $Z$ of the target and on the curvature radius
$R$ have been investigated.
Both crystals have the diamond cubic crystal structure but differ
significantly in the atomic number: $Z_{\rm Si}=14$ vs.
$Z_{\rm Ge}=32$.

The parameters used in the simulations (these include crystals'
thickness, orientation with respect to the incident beam,
curvature radii)
have been chosen in accordance with the experiments
\cite{SytovEtAl:EPJC_v77_901_2017,BandieraEtAl:EPJC_v81_284_2021}
performed at the Mainz Mikrotron (MAMI) facility.
The thickness along the incident beam direction was
$L=15$ microns, which is comparable with the dechanneling length
of 855 MeV electrons
\cite{SytovEtAl:EPJC_v77_901_2017,
KorolSushkoSolovyov:EPJD_v75_p107_2021,
SushkoEtAl:JPConfSer_v438_012019_2013,
SushkoEtAl:JPConfSer_v438_012018_2013,
PolozkovEtAl:EPJD_v68_268_2014,
Sushko:Thesis_2015},
 thus increasing the efficiency of the orientational effects.

\begin{figure*}[h]
\centering
\includegraphics[width=1\linewidth]{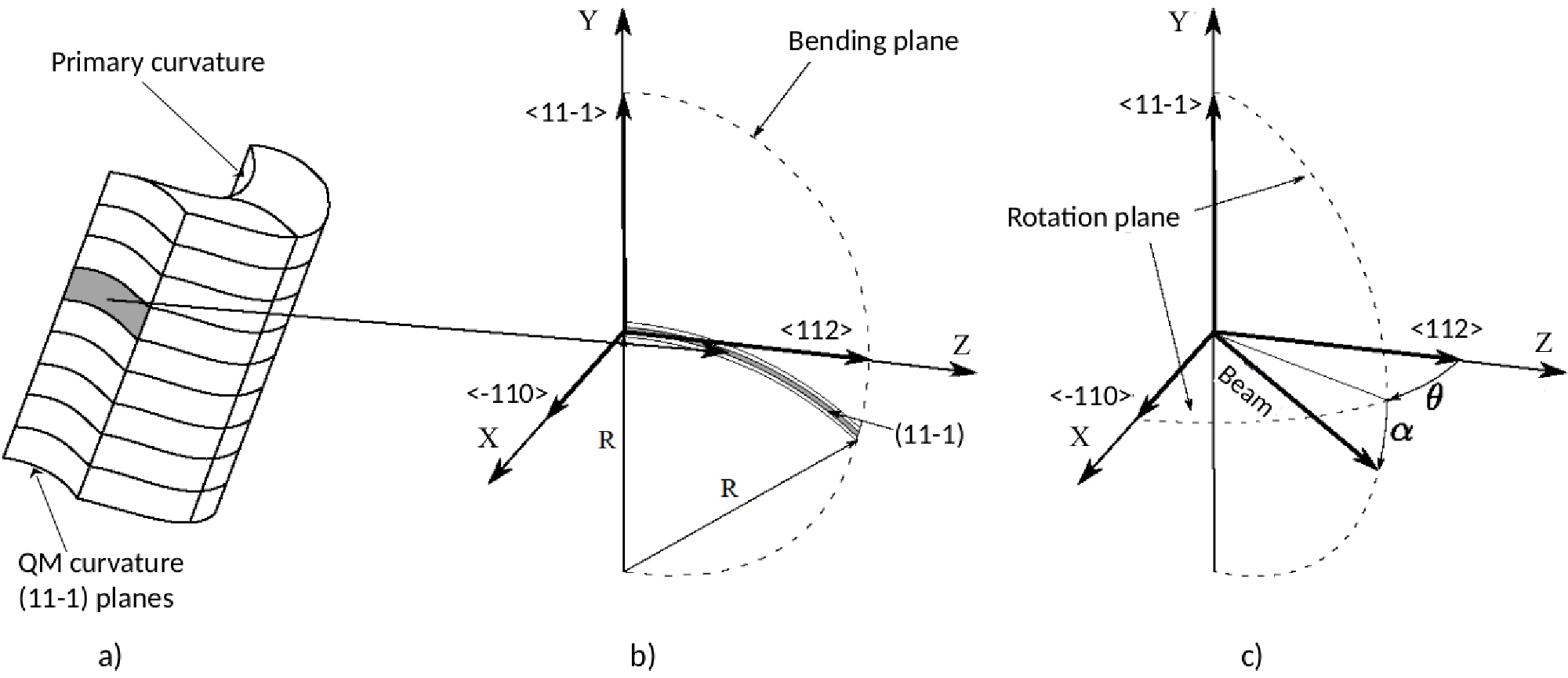}
\caption{
Crystal geometry and its orientation with respect to the incident
beam that was used in the experimental setup and in the current
simulations.
\textit{(a)} The $(11\bar{1})$ planes (the shaded area marks one
of these) of the crystal experience the quasi-mosaic (QM)
secondary bending as a result of primary curvature due to
mechanical bending.
\textit{(b)}
Crystallographic axes chosen for the Cartesian coordinate system.
The $(XZ)$ plane corresponds to the unbent $(11\bar{1})$
crystallographic plane.
Shaded strip illustrates the QM bending; $R$ denotes the radius of
the QM curvature.
\textit{(c)}
Angles used to characterize the incident electron beam velocity
$\bfv_0$ relative to the crystal orientation:
$\theta$ is the angle between $\bfv_0$ and the
$\langle 112\rangle$ axial direction,
$\alpha$ -- the angle between $\bfv_0$ and the unbent
$(11\bar{1})$ plane.
}
\label{Figure01.fig}%
\end{figure*}

\section{Beam-crystal alignment \label{Alignment}}

In the experiments
\cite{SytovEtAl:EPJC_v77_901_2017,BandieraEtAl:EPJC_v81_284_2021},
a uniform bending of $(11\bar{1})$ crystalline planes was achieved
by means of a specially designed mechanical holder
\cite{SalvadorEtAl:JINST_v13_C04006_2018},
which allowed one remotely alter the curvature radius thus making
possible to investigate the deflection efficiency and the
radiation intensity as functions of $R$.
Bending of the planes occurs due to a secondary
‘quasi-mosaic’ (QM) deformation
\cite{IvanovEtAl:JETPLett_v81_p977_2005,
GuidiEtAl:JPD_v42_182005_2009,CamattariEtAl:JAC_v489_p977_2015}
that was the result of the primary deformation caused by the
holder, see illustrative Fig. \ref{Figure01.fig}a.
Prior to performing the beam experiment,
the orientation of crystallographic directions was determined by
means of a high-resolution X-ray diffraction
\cite{GuidiEtAl:JPD_v42_182005_2009},
see Fig. \ref{Figure01.fig}b.
This technique was also used to assess the uniformity of the
crystal bending at different curvatures.
Mounting crystals on a goniometer with three degrees
of freedom provided high precision in the angular alignments
between the samples and the incident beam
\cite{BackeEtAl:NIMB_v266_p3835_2008}.

Following the experimental setup, we have simulated passage
of the electrons and calculated the emission spectra for the
following two different beam--crystal alignments that can be
illustrated by Fig. \ref{Figure01.fig}c:
 \begin{enumerate}
\item[(i)] \textit{Planar channeling alignment.}
In this case, the beam velocity, being aligned at the entrance
with the  $(11\bar{1})$ plane, is directed at the angle $\theta$
with respect to the $\langle 112\rangle$ axis.
 To avoid axial channelling, $\theta$ must to be chosen much
 larger than Lindhard’s critical angle for the
$\langle 112\rangle$ axis (ca 0.4 and 0.6 mrad
for Si and Ge crystals, respectively) but much smaller than
190 mrad that is the angle between the
$\langle 112\rangle$ and $\langle 123\rangle$ axes
\cite{HaurylavetsEtAl:EPJPlus_v137_34_2022}.
In the current simulations the value $\theta=95$  mrad was used.
The second angle $\alpha$, indicated in Fig. \ref{Figure01.fig}c,
was set to zero.

\item[(ii)] \textit{Volume reflection alignment} corresponds
to the incident beam angle $\alpha$ within the range
$\theta_{\rm L} \lesssim \alpha < L/R$ where
$\theta_{\rm L}=(2U_0/\E)^{1/2}$
denotes Lindhard’s critical angle for the planar channeling
along the $(11\bar{1})$ planes.
As a result, at the entrance most of the particles move
in the over-barrier mode across the bent channels but later
on in the bulk they can experience either volume capture
or volume reflection \cite{TaratinVorobiev:PLA_v115_p398_1986,%
TaratinVorobiev:PLA_v119_p425_1987}.

 \end{enumerate}

Within the continuous potential approximation Lindhard’s critical
angle is given by  $\theta_{\rm L}=(2U_0/\E)^{1/2}$
with $U_0$ standing for the depth of the continuous
interplanar potential.
Using the values $U_0\approx 24$ eV for the Si$(11\bar{1})$
channel and $U_0\approx 32$ eV for the Ge$(11\bar{1})$
channel, calculated within the Pacios approximation
(see Fig. \ref{Appendix:fig.03}
in Section \ref{Appendix}),
one obtains $\theta_{\rm L}\approx 0.24$ and 0.27 mrad
for a 855 MeV electron channeling in the silicon and germanium
crystals, respectively.

For each set $(\theta, \alpha)$ considered in the simulations,
to build the angular distribution of deflected electrons
a large number, from $N\approx 4\times 10^4$ up to
$15\times 10^4$,
of trajectories have been simulated and analyzed.
The computation of spectral distributions of the emitted radiation
has been carried out using lower number of the trajectories,
$N\gtrsim 10^4$.
The analysis was performed for different values of the crystals'
curvature.

To calculate the angular distributions, the trajectories have been
simulated using the Moli\'{e}re and the Pacios potentials for the
electron--atom interaction.
The analysis performed in Ref.
\cite{HaurylavetsEtAl:EPJPlus_v137_34_2022}
has shown that the Pacios approximation provides better agreement
with the experiment for the spectrum of channeling radiation.
Therefore, in the current paper the spectral dependencies have
been obtained using this potential only.

Below in the paper, the statistical errors indicated for the
simulated data are due to the finite number of the trajectories.
The experimental data shown have been obtained by
digitalizing the graphical data presented in Refs.
\cite{SytovEtAl:EPJC_v77_901_2017,BandieraEtAl:EPJC_v81_284_2021}.
Finite width of the lines used to draw experimentally measured
angular distributions corresponds to the statistical errors
of the data.
The error bars for the experimental data on the emission spectra
have not been indicated in the cited papers.

\section{Case studies \label{CaseStudies}}

\subsection{Angular distribution of deflected electrons
\label{AngDistr}}

Bending curvature of crystalline targets used in the experiment
\cite{SytovEtAl:EPJC_v77_901_2017} was characterized
in terms of the bending angle, $\theta_{\rm b} = L/R$.
For the crystals probed the quoted values of $\theta_{\rm b}$
are $315, 550, 750, 1080$ $\mu$rad and
$820, 1200, 1430$ $\mu$rad for the $L=15$ $\mu$m thick
Si and Ge crystals, respectively.
Below in this section the curvature is specified in terms of
the curvature radius,
the corresponding values of which are as follows:
$R=47.6, 27.3, 20.0, 13.9$ mm for Si and
$R=18.3, 12.5, 10.5$ mm for Ge.

Distributions (in mrad$^{-1}$) shown below in this Section have
been normalized to the unit area within the intervals
of the deflection angle  $\vartheta$ (in mrad) indicated in Ref.
\cite{SytovEtAl:EPJC_v77_901_2017}.

The deflection angle stands for the projection of the angle
between initial $\bfv_0$ and final $\bfv$ velocities of the
projectile electron on the plane that contains
$\bfv_0$ and the $\langle 11\bar{1}\rangle$ axis.

\subsubsection{Channeling alignment \label{AngDistr:ChanGeom}}

\begin{figure}[h]
\centering
\includegraphics[clip,width=12cm]{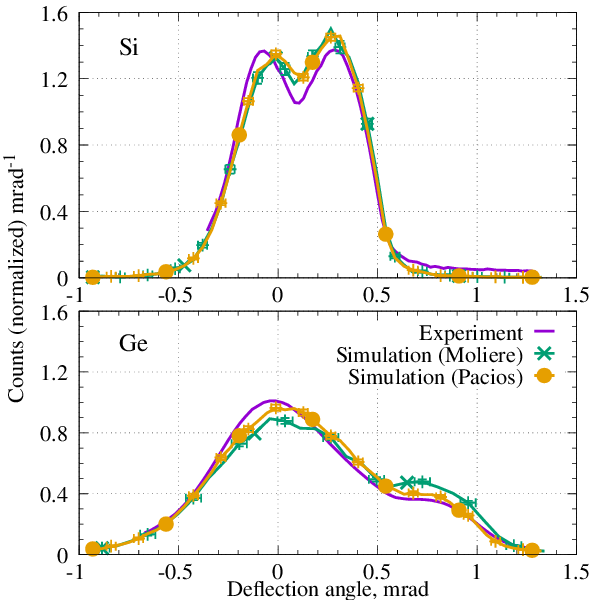}
\caption{
Simulated versus experimentally measured (see Ref.
\cite{SytovEtAl:EPJC_v77_901_2017}) angular distributions of
deflected electrons in the case of the channeling alignment
of the incident beam with the $(11\bar{1})$  plane.
The incident angles are $\theta=95$ mrad, $\alpha=0$, see Fig.
\ref{Figure01.fig}.
The simulations has been carried out with both the Moli\'{e}re and
Pacios atomic potentials.
The curvature radius $R$ of the plane equals to 47.6 mm
($\theta_{\rm b}=0.315$ mrad) for the
silicon crystal target (top) and to 18.3 mm
($\theta_{\rm b}=0.820$ mrad) for
the germanium crystal (bottom).
}
\label{Figure02.fig}
\end{figure}

Figure \ref{Figure02.fig} presents angular distributions of
electrons after passing through bent silicon
(top graph,  $R=47.6$ mm, $\theta_{\rm b}=0.315$ mrad) and
germanium (bottom graph, $R=18.3$ mm, $\theta_{\rm b}=0.820$ mrad)
crystals.
The incident particle's velocity $\bfv_0$ is directed along the
$(11\bar{1})$ crystallographic plane.

Two maxima are distinguishable in the curves presented.
The left one, located in the vicinity of $\vartheta=0$,
is mainly due to the electrons that experience the
over-barrier motion at the crystal entrance.
The right maximum,  centered at $\vartheta = \theta_{\rm b}$,
corresponds to the projectiles that move in the channeling mode
at the crystal exit.
Its width equals to Lindhard's critical angle $\theta_{\rm L}$.

On average, the scattering angle in electron collisions with
heavier germanium atom is larger than with a silicon atom.
This relation also holds for the multiple scattering angles
in germanium and silicon media.
As a result,
(i) the over-barrier peak in the distribution for the silicon
target is higher and narrower than for the germanium target,
(ii) the dechanneling rate in Ge crystal is notably
larger than in Si, so that the channeling peak in the
bottom graph in Fig. \ref{Figure02.fig} is much less intensive
than in the top graph.
Analysis of the trajectories simulated in these
case studies has shown that approximately 30 \% of the
electrons incident on the Si target pass through the  crystal
moving in the channeling mode whereas for the Ge target this
number is much lower, 5 \%.

For the silicon target, a discrepancy between the
distributions calculated using the Moli\'{e}re and Pacios atomic
potentials are within the statistical uncertainties whereas
for the germanium crystal
the difference between the distributions is clearly seen in the
vicinities of both maxima.
For the heavier crystal, the Pacios approximation, being more
accurate in description of the atomic potential (see Appendix
\ref{Appendix}), results in a better agreement of the simulated
angular distribution with the experimentally measured one.

\begin{figure}[h]
\centering
\includegraphics[clip,width=12cm]{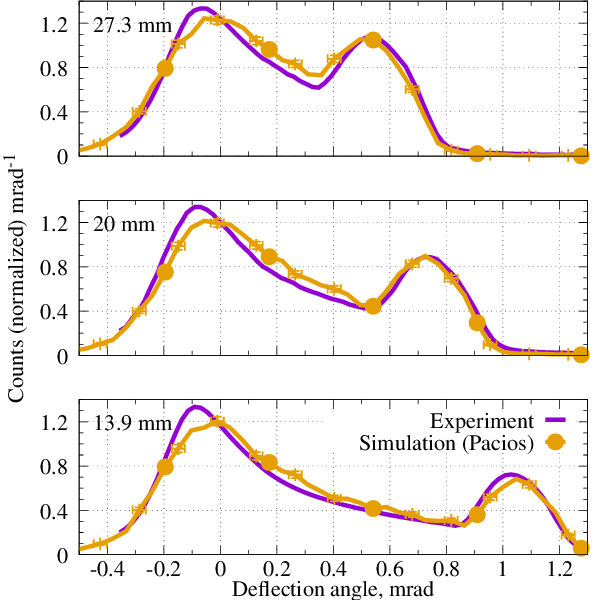}
\caption{
Angular distribution of deflected electrons after
interaction with bent Si crystal as a function of the
curvature radius $R$ of the $(11\bar{1})$ plane: $R=27.3$ mm
(top), 20 mm (middle) and 13.9 mm (bottom).
The incident beam geometry corresponds to the planar channeling
alignment with  $\theta=95$ mrad and $\alpha=0$.
The experimental data are from Refs.
\cite{SytovEtAl:EPJC_v77_901_2017,%
HaurylavetsEtAl:EPJPlus_v137_34_2022}.
The results of the simulations shown correspond to the Pacios
atomic potential.
}
\label{Figure03.fig}
\end{figure}

Evolution of the angular distribution with decrease in the
curvature radius (increase in the bending angle) of the silicon
crystal is demonstrated by Fig. \ref{Figure03.fig}.
The values of $R$ indicated in the top, middle and bottom graphs
correspond to $\theta_{\rm b}= 0.55, 0.75, 1.08$ mrad.
It is seen that the position of the channeling peak in the
simulated distribution follows the quoted $\theta_{\rm b}$
values and its position, height and width agree well with the
experimental data.
In all simulated distributions the position and height of the
maximum due to the over-barrier particles at the entrance
are the same within the statistical errors.
There are systematic deviations from the experiment.
First, the simulation underestimates (by ca 10 \%)
the peak values of the maxima.
Second, the experimentally measured distributions are peaked at
$\vartheta\approx -0.08$ mrad whereas the
maxima in the simulated distributions are less shifted
to the domain of negative scattering values.
Finally, the simulation produces higher values in the region between
both peaks which corresponds to the higher dechanneling rate.

\begin{figure}[h]
\centering
\includegraphics[clip,width=12cm]{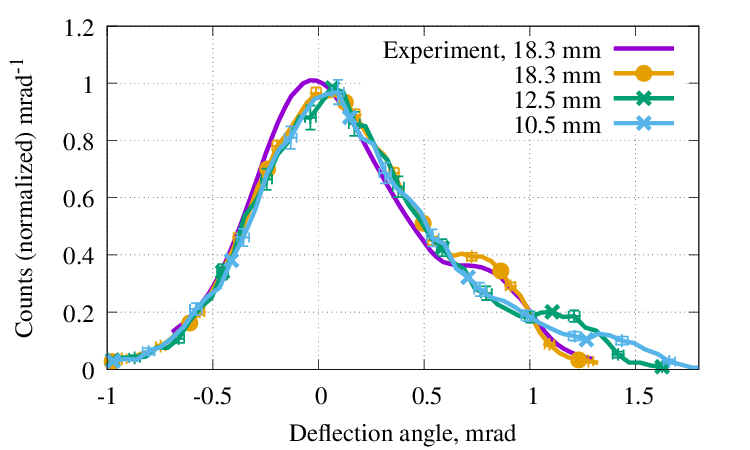}
\caption{
Angular distribution of deflected electrons after
interaction with bent Ge crystal as a function of the
curvature radius $R$ of the $(11\bar{1})$ plane as indicated
in the legend.
The incident beam geometry corresponds to the planar channeling
alignment with  $\theta=95$ mrad and $\alpha=0$.
The experimental data for $R=18.3$ mm  are from Ref.
\cite{SytovEtAl:EPJC_v77_901_2017}.
The results of the simulations shown correspond to the Pacios
atomic potential.
}
\label{Figure04.fig}
\end{figure}

Angular distribution of the electrons deflected by germanium
crystals bent with different curvature radii are shown in
Fig. \ref{Figure04.fig}.
In Ref. \cite{SytovEtAl:EPJC_v77_901_2017} the experimental data
are presented only for $R= 18.3$ mm ($\theta_{\rm b} =0.82$ mrad).
As in the case of the silicon crystal, the position of the
channeling maximum is shifted towards higher deflection angles as
the bending curvature $1/R$ increases.
Its peak value becomes less intensive due to higher dechanneling
rates for larger  curvatures.

For all curvature radii considered the distributions of the
over-barrier electrons at the entrance obtained in the simulations
are centered at $\vartheta\approx 0$.
The experimentally measured distribution exhibits some shift
towards negative values of the deflection angle.

In Ref. \cite{SytovEtAl:EPJC_v77_901_2017} the fitting procedure
is described which has been used to extract the dechanneling
length $\Ld$ and channeling efficiency $\eta$ from the angular
distributions.
The channeling efficiency has been defined
as the integral value of the gaussian fit of the channeling
peak carried out over the interval $\pm$ three standard
deviations from the peak's position.

Within the relativistic molecular dynamics framework utilized in
this paper, these quantities can be calculated directly
by means of statistical analysis of the trajectories
(see Ref. \cite{KorolSushkoSolovyov:EPJD_v75_p107_2021} for more
details).
To determine the average interval within which a particle moves
in the channeling mode starting at the crystal entrance one
can analyse the trajectories of the accepted particles.
An accepted projectile, stays in the channeling mode of motion
over some interval until an event of the dechanneling
(if it happens within the crystal thickness considered).
Hence, the dechanneling effect for the accepted particles can
be characterized in terms of the penetration length $\Lp$
\cite{MBN_ChannelingPaper_2013}
defined as the arithmetic mean of the initial channeling segments
$L_{\rm ch0}$ calculated with respect to all accepted
trajectories:
$\Lp  =
N_{\rm acc}^{-1}\sum_{j=1}^{N_{\rm acc}} L_{\rm ch0}^{(j)}$.
The channeling efficiency is defined through the
ratio $\eta = N_{\rm ch}/N$ where $N$ is the number of
the incident electrons and $N_{\rm ch}$ stands for the number of
electrons that leave the crystal in the channeling mode
(this number accounts for the projectiles that channel through the
whole crystal as well as for those that were captured in the
channeling mode somewhere in the bulk and channel till the
crystal exit).

\begin{table}[h]
\caption{
Curvature radius, $R$, bending angle, $\theta_{\rm b}$,
penetration length $\Lp$ and
channeling efficiency $\eta$ for silicon and germanium
crystals.
Three last columns present experimentally measured
dechanneling length, $L_{\rm d, exp}$, and
channeling efficiency $\eta_{\rm exp}$
and the simulated value of channeling efficiency
\cite{SytovEtAl:EPJC_v77_901_2017}.
}
\begin{tabular*}{\textwidth}{@{}crrrrrrrr}
\hline
\hline
  &$R$  &$\tb$ & $\Lp$       &$\eta$         &$L_{\d,\exp}$& $\eta_{\exp}$
                                                           & $\eta_{\rm sim}$\\
  &(mm) &(mrad)&($\mu$m)&  & ($\mu$m) &\\
\hline
  &27.3 & 0.55 &15.7$\pm$0.3 &0.30$\pm$0.02  & 17.7$\pm$3.0& 0.248$\pm$0.016
                                                           & 0.3000 \\
Si&20.0 & 0.75 &14.8$\pm$0.4 &0.24$\pm$0.02  & 14.0$\pm$2.2& 0.206$\pm$0.013
                                                           & 0.2519 \\
  &13.9 & 1.08 &13.1$\pm$0.3 &0.18$\pm$0.01  & 10.1$\pm$1.0& 0.165$\pm$0.010
                                                           & 0.1907  \\
\hline
  &18.3 & 0.82 & 7.3$\pm$0.2 &0.110$\pm$0.010&  9$\pm$5    & 0.084$\pm$0.017
                                                           & 0.0909  \\
Ge&12.5 & 1.20 & 6.8$\pm$0.1 &0.058$\pm$0.011& 7.3$\pm$1.2 & 0.036$\pm$0.007
                                                           & 0.0468\\
  &10.5 & 1.43 & 6.4$\pm$0.2 &0.037$\pm$0.011& 5.9$\pm$1.5 & 0.019$\pm$0.004
                                                           & 0.0320\\
\hline
\hline
\end{tabular*}
\label{Table.01}
\end{table}

Table \ref{Table.01} summarizes the values of $\Lp$ and $\eta$
that have been calculated for silicon and germanium crystals with
the bending parameters as indicated.
Also presented are the data from Ref.
\cite{SytovEtAl:EPJC_v77_901_2017}
(see tables 1-3 there): experimentally measured dechanneling
lengths and channeling efficiency as well as the latter
quantity, notated as
$\eta_{\rm sim}$,  obtained by means of the
CRYSTAL code \cite{SytovTikhomirov_NIMB_v355_p383_2015}
(we quote the $\eta_{\rm sim}$ values
without the statistical uncertainties which are
within the range $(2-4)\times10^{-4}$).

It is seen that with account for the statistical errors our
results for $\Lp$ fully correlate with the experimental data
for $\Ld$ (as well as with the results obtained with the
CRYSTAL that is not  shown here but can be found in table 3
in Ref.  \cite{SytovEtAl:EPJC_v77_901_2017} ).

For the germanium crystals the values of
channeling efficiency obtained in the current simulations are
systematically higher than measured experimentally.
We note, however, that similar tendency is seen
in $\eta_{\rm sim}$,
see the last column in Table \ref{Table.01}.

\subsubsection{Volume reflection alignment
\label{AngDistr:VR}}

In second case study, the geometry of which is illustrated
in  \ref{Figure01.fig}c, the beam was incident at the angle
$\alpha=0.45$ mrad with respect to the $(11\bar{1})$ plane,
which is larger than Lindhard's critical angles in Si and Ge
crystals.
Hence, at the crystal entrance the electrons are not accepted
into the channeling mode of motion.
In this case, due to the bending of crystalline planes that
results in the asymmetry of the interplanar potential,
a particle can be captured into the channeling mode somewhere
in the crystal volume (volume capture) or be reflected from the
interplanar potential barrier (volume reflection).
The volume capture leads to the appearance of the channeling peak
in the angular distribution whereas the volume reflection
shifts the over-barrier peak to the domain of negative deflection
angles.

\begin{figure}[h]
\centering
\includegraphics[clip,width=12cm]{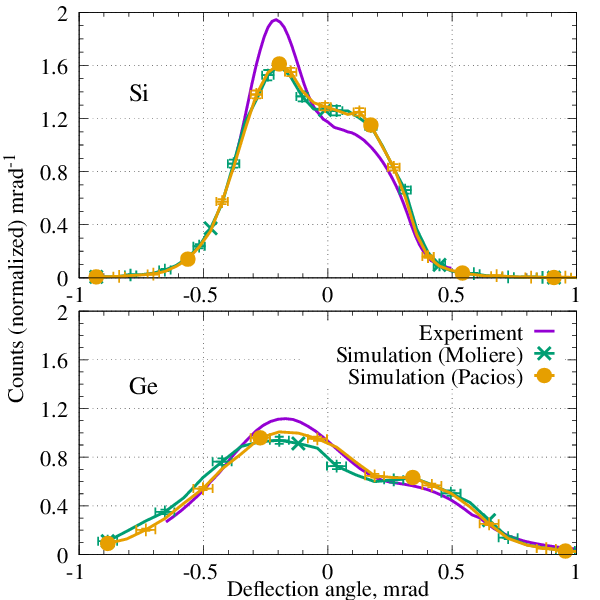}
\caption{
Angular distribution of deflected electrons after interaction with
Si$(11\bar{1})$ crystal (top)
bent with $R=47.6$ mm
($\theta_{\rm b}=0.315$ mrad)
 and Ge$(11\bar{1})$ crystal (bottom) bent with $R=18.3$ mm
 ($\theta_{\rm b}=0.820$ mrad).
The experimentally measured \cite{SytovEtAl:EPJC_v77_901_2017} and
the simulated distributions correspond to the volume reflection
alignment with $\alpha=0.45$ mrad and $\theta=95$ mrad, see
Fig. \ref{Figure01.fig}c.
}
\label{Figure05.fig}
\end{figure}

Figure \ref{Figure05.fig} compares the results of the simulations
carried out for bent silicon
($R=47.6$ mm, $\theta_{\rm b}=0.315$ mrad) and
germanium ($R=18.3$ mm, $\theta_{\rm b}=0.820$ mrad)
crystals with the experimental data from
Ref. \cite{SytovEtAl:EPJC_v77_901_2017}.

For comparison, the simulations have been performed utilizing
each one of the atomic potentials.
As in the case of the channeling geometry,
Fig. \ref{Figure02.fig}, both atomic potentials produce similar
(within the statistical error)
angular distributions for the silicon target whereas
for the germanium crystal the difference between them
is clearly seen.

The simulations follow the experiment in reproducing the
changes in the angular distributions as compared to the
channeling alignment.
To be noted is the shift, by approximately -0.2 mrad
for both crystals, of the over-barrier peak caused by the
volume-deflected electrons.
The contribution of these electrons explains the increase
in the peaks' values in comparison with the over-barrier
maxima in Fig. \ref{Figure02.fig}.

The channeling peaks, however, are modified differently
for the silicon and germanium targets both in terms of
the shift and the peak value.
Let us first consider the Ge crystal,
Fig. \ref{Figure05.fig} bottom.
The incident angle $\alpha=0.45$ mrad is smaller
than the bending angle $\theta_{\rm b}=0.820$ mrad, therefore,
geometrically, the line drawn along the beam direction $\bfv_0$
becomes tangent to the bent plane at the distance
$l=L(\theta_{\rm b}-\alpha)/\theta_{\rm b}\approx 7$ $\mu$m
from the crystal exit.
Since $l$ is ca 2 times less than the crystal thickness,
a particle that experiences volume capture at this distance
has much higher probability to exit in the channeling
mode than the same particle accepted at the entrance in the
channeling geometry.
These qualitative arguments explain the increase in the
channeling  peak intensity for Ge in Fig. \ref{Figure05.fig} in
comparison to that in Fig. \ref{Figure02.fig}, as well as the
shift of the position: instead of  0.820 mrad the maximum is at
ca $\theta_{\rm b}-\alpha = 0.37$ mrad.
For the silicon crystal, Fig. \ref{Figure05.fig} top,
the geometrical approach has to be modified since in this case
$\alpha>\theta_{\rm b}=0.315$ mrad so that the line $\bfv_0$ does
not touch the curved channel.
Assuming that a particle can be accepted into the channeling mode
if its local incident angle with respect to the plane is less
than Lindhard's critical, one estimates that
some fraction of electrons can experience volume capture
at the distances
$l=L(\theta_{\rm L}+\theta_{\rm b}-\alpha)/\theta_{\rm b}\approx 5$
$\mu$m.
The probability for an electron to channel over such distance is
high enough, however, the number of volume capture events is
smaller than the number of the accepted particles in the
case of the channeling alignment.
As a result, the channeling peak in Fig. \ref{Figure05.fig} top
is lower than that in Fig. \ref{Figure02.fig} top.

Apart from the aforementioned similarities, there are
differences in the distributions' maxima values measured
experimentally and obtained via the simulations.
The differences are more pronounced for the Si target:
the experimental over-barrier peak is ca 20\% higher whereas the
channeling peak is lower by approximately the same amount.
For the Ge crystal the simulation-to-experiment
agreement is better if one considers the curve corresponding to
the Pacios potential: the maxima differ by 10\%.

\subsection{Radiation spectra \label{Spectra}}

Characterization of the radiation emitted by the beam passing
though a
target can be done in terms of spectral distribution of the
radiant energy, $E \d N /\d E$.
Here $\d N$ stands for the number of photons emitted within the
energy interval $[E,E+\d E]$.
To compute the spectral distribution, the trajectories simulated
with the use of the Pacios potential have been utilized further
following the formalism and algorithm
described in detail in Ref. \cite{MBN_ChannelingPaper_2013}.
To obtain $E \d N /\d E$ per particle, first the
spectral-angular distribution was calculated for each
trajectory, then it was integrated over the cone
$\theta_{0} = 4.65$ mrad along the incident beam, and the
results were averaged over all trajectories.
The quoted value of $\theta_{0}$ is approximately eight times
larger than the natural emission cone $\gamma^{-1}$ of
a 855 MeV electron and is also much larger than the
bending angles $\theta_{\rm b}=L/R$
(see below).
Therefore, the calculated spectra accounted practically
for all radiation emitted by the projectile particles.

The spectral distributions have been calculated for the planar
channeling and volume reflection alignments.
Both cases refer to the motion in an oriented crystalline
target where a projectile experiences collective action of the
electrostatic
fields of the atoms.
This leads to the increase in the radiant energy as compared to
that
emitted in the corresponding amorphous medium via the incoherent
bremsstrahlung process.
To quantify the latter process, the trajectories of 855 MeV
electrons passing through 15 microns thick amorphous silicon
and germanium targets have been simulated followed by the
calculation of the spectral distributions as described above.
For the sake of reference, the values of $E \d N /\d E$
calculated for the amorphous media are presented below:
\begin{eqnarray}
\hspace*{-1cm}
\left. E{\d N \over \d E}\right|_{\mathrm{Si, am}}
=
\left(0.21\pm 0.02\right)\times10^{-3},
\qquad
\left.E {\d N \over \d E}\right|_{\mathrm{Ge, am}}
=
\left(0.85\pm 0.05\right)\times10^{-3}
\,.
\label{Spectra:eq.01}
\end{eqnarray}
These values, that characterize the incoherent bremsstrahlung
background within the photon energy range $E=1 - 10$ MeV,
are consistent with the data measured experimentally,
see figure 2 in Ref. \cite{BandieraEtAl:EPJC_v81_284_2021}.
The intensity of radiation in the amorphous germanium environment
is approximately four times higher than in the silicon one.
This factor correlates with the ratio of atomic numbers squared.

In the cited paper the curvature of crystals was quantified in
terms of the bending angle $\theta_{\rm b}$ equal to
$550, 750, 1080$ $\mu$rad
for the silicon crystal and to $820, 1200, 1430$ $\mu$rad
for the germanium crystal.
For the thickness $L=15$ $\mu$m these values produce the curvature
radii (in mm)
equal to $27.3, 20.0, 13.9$ and $18.3,12.5,10.5$, correspondingly.

\subsubsection{Channeling alignment \label{Spectra:ChanGeom}}

\begin{figure}[h]
\centering
\includegraphics[clip,width=12cm]{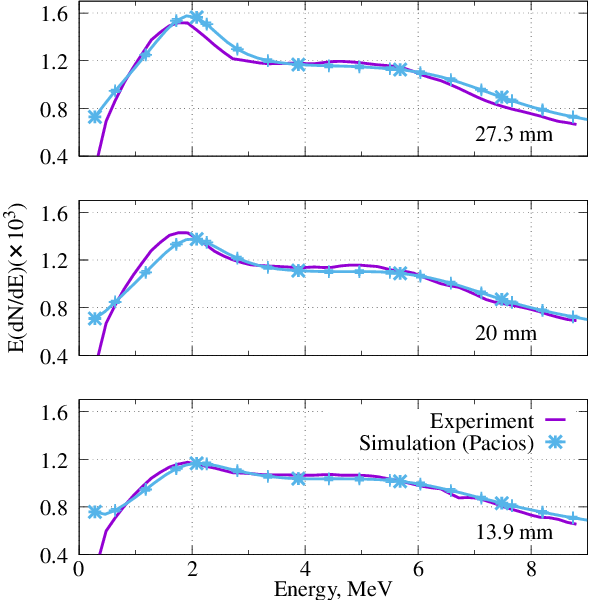}
\caption{
Spectral distribution of radiation emitted by 855 MeV electrons
passing through Si$(11\bar{1})$ crystal bent with the
curvature radius 27.3 mm (top), 20 mm (middle) and 13.9 mm (bottom).
The incident beam geometry corresponds to the planar channeling
alignment with  $\theta=95$ mrad and $\alpha=0$.
Experimental data are taken from \cite{BandieraEtAl:EPJC_v81_284_2021}.
Note that the values of $E \d N /\d E$ shown are multiplied
by the factor $10^3$.
}
\label{Figure06.fig}
\end{figure}

\begin{figure}[h]
\centering
\includegraphics[clip,width=12cm]{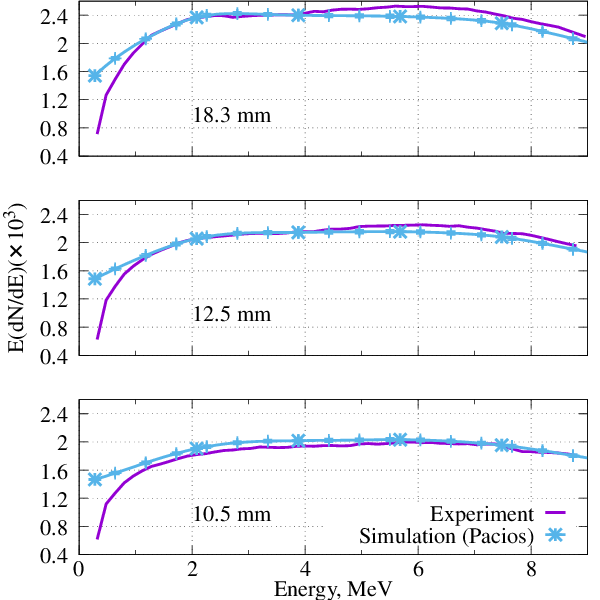}
\caption{
Same as in Fig. \ref{Figure06.fig} but for Ge$(11\bar{1})$
crystal.
The curvature radii are indicated in each graph.
}
\label{Figure07.fig}
\end{figure}

For the channeling alignment,
comparison of spectral distributions measured
experimentally \cite{BandieraEtAl:EPJC_v81_284_2021} and
calculated within the current simulations is presented by Fig.
\ref{Figure06.fig} (silicon) and Fig. \ref{Figure07.fig}
(germanium).
One can note an overall good agreement between the
experiment and the theory.

A common feature, seen for both crystals and all curvature radii,
is an increase in the radiation intensity in the oriented
targets in comparison to its values in the amorphous media,
Eq. (\ref{Spectra:eq.01}).
For Si crystal bent with $R=27.3$ mm the enhancement factor is
about eight at the photon energy $E\approx 2$ MeV .
For the Ge crystal with $R=18.3$ mm it stays equal to
approximately three over a wide range of the photon energies.

The enhancement is mainly due to the emission by the particles
that move in the channeling mode of motion.
As the bending curvature $1/R$ increases, the average length of
the channeling segment, $\langle L_{\rm ch} \rangle$,
decreases resulting in the decrease in the radiant energy which is
proportional to $\langle L_{\rm ch} \rangle$.
For silicon crystals with $R=27.3$ and 13.9 mm,
statistical analysis of the lengths of all channeling segments
in all trajectories simulated has led to
$\langle L_{\rm ch} \rangle=7.5 \pm 0.1$ and
$5.8 \pm 0.1$ $\mu$m, respectively.
The ratio of the central values, $7.5/5.8\approx 1.3$, nicely
correlates with the height of the peak at $E\approx 2$ MeV
in the top
($E \d N /\d E\approx 1.6\times 10^{-3}$) compared to that in the
bottom ($E \d N /\d E\approx 1.2\times 10^{-3}$) graphs in
Fig. \ref{Figure06.fig}.
Similar calculations performed for the germanium targets
resulted in $\langle L_{\rm ch} \rangle=4.8 \pm 0.1$ and
$3.5 \pm 0.1$ $\mu$m for $R=18.3$ and $10.5$ mm, which also
corresponds to the ratio of the intensities in the top and bottom
graphs in Fig. \ref{Figure07.fig}.

As in the case of amorphous environment, the intensity in oriented
germanium crystal is higher that in the silicon crystal, although
the excess is not as large as for the amorphous targets,
see Eq. (\ref{Spectra:eq.01}).
Partly, this change is due to the  excess of
$\langle L_{\rm ch} \rangle_{\rm Si}$ over
$\langle L_{\rm ch} \rangle_{\rm Ge}$.

A feature, which distinguishes the spectra in Si and Ge,
refers to the peak of channeling radiation seen
in the Si spectra at $E\approx 2$ MeV but not pronounced for Ge.
An initial intuitive explanation could be as follows.
To a great extent, the peaks are due to the
radiation emitted by electrons that move in the channeling
mode through the whole crystal.
For the silicon crystal the fraction of such particles
with respect to the number of incident particles
is quite large, being approximately equal to
25\%, 20\%, 16\%
for the $R$ values indicated in Fig. \ref{Figure06.fig}
from top to bottom.
For the germanium target the fraction is nearly
an order of magnitude smaller, equal to 3.6\%, 2.5\% and
1.9\% with reference to Fig. \ref{Figure07.fig}.
In this case, main contribution to the spectrum comes from the
over-barrier projectiles, which radiate in a broader energy
range and, thus, level the spectrum's shape.

\begin{figure}[h]
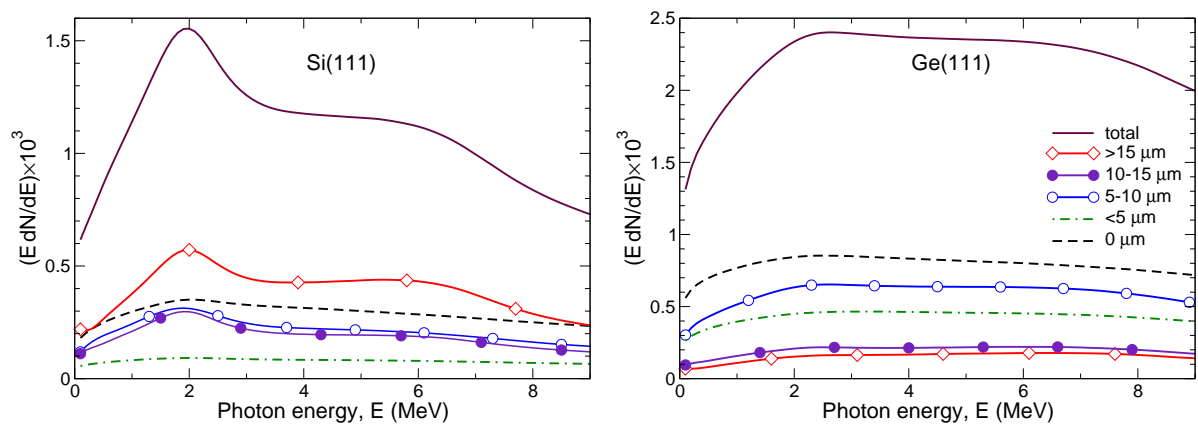

\centering
\includegraphics[clip,width=7.75cm]{Figure08a.eps}%
\hspace*{0.2cm}
\includegraphics[clip,width=7.75cm]{Figure08b.eps}
\caption{
Spectral distribution of radiation in Si
($R=27.3$ mm, left graph) and
Ge ($18.3$ mm, right graph) bent crystals with explicit contributions from different group of particles as indicated
in the common legend in the right graph.
See explanations in the text.
The incident beam geometry corresponds to the planar channeling
alignment.
}
\label{Figure08.fig}
\end{figure}

To check this hypothesis, one can analyse the contributions
of different groups of particles to the total spectrum.
The result of this analysis is presented in Fig.
\ref{Figure08.fig} for crystals bent
with the largest curvature radii:
Si with $R=27.3$ mm (left) an Ge  with $R=18.3$ mm (right).
In both graphs solid curve shows the total spectrum
(per particle);
solid line + diamonds stands for the contribution from
the electrons that move in the channeling mode
through the whole crystal;
solid line + closed circles, solid line + open circles and
dash-dotted line denote the contribution from the
electrons with the initial channeling segment
$L_{\rm ch0}$ (in microns) lying within the interval:
$10\leq L_{\rm ch0} <15$, $5\leq L_{\rm ch0} <10$,
and $0\leq L_{\rm ch0} <5$, respectively.
The contribution from the trajectories that have no initial
channeling segments (but which can have those in the bulk)
is shown with dashed line.
For Si, the main channeling peak is indeed due to the particles
that channel through the whole crystal
(30\% of the peak intensity).
The emission spectra of the particles
that channel over considerable part of the crystal (the curves
with the circles of both types) are also enhanced at
$E\approx 2$ MeV.
The emission of other particles provide a smooth background.
In contrast, with respect to the shape of the spectrum, all
groups of particles passing through the germanium crystal,
including those which move in the channeling mode
radiate similarly: a smooth curve with nearly constant value within the interval
$E=2-8$ MeV.

\begin{figure}[h]
\centering
\includegraphics[clip,width=12cm]{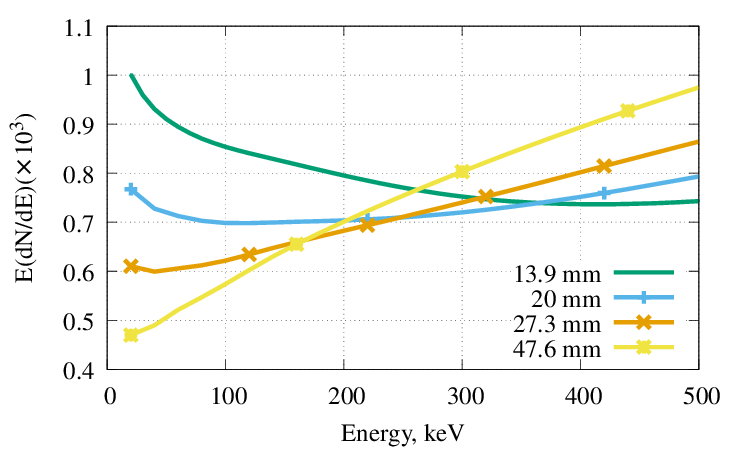}
\caption{
Low-energy part of the radiation spectra by 855 MeV electrons
in bent Si crystal calculated for different curvature radii $R$
as indicated.
The increase in the spectra as the photon energy decreases,
seen for all radii but $R=47.6$ mm, is due to the synchrotron
radiation.
The incident beam geometry corresponds to the planar channeling
alignment with  $\theta=95$ mrad and $\alpha=0$.
}
\label{Figure09.fig}
\end{figure}

An additional feature that appears in the emission spectrum in
a bent crystal as compared to a linear crystal is due to a
circular motion of the channeling particles.
This motion gives rise to the synchrotron-type radiation,
which contributes to the low-energy part of the spectrum.
This feature was predicted theoretically long ago
\cite{KaplinVorobiev:PLA_v67_p135_1978,
TaratinVorobiev:NIMB_v31_p551_1988}
and has also been analysed quantitatively by means of
all-atom relativistic molecular dynamics simulations
\cite{SushkoEtAl:JPConfSer_v438_012018_2013,
PolozkovEtAl:EPJD_v68_268_2014,ShenEtAl:NIMB_v424_p26_2018}.
The characteristic energy $E_{\rm c}$ beyond which the
synchrotron radiation intensity rapidly falls off
\cite{Jackson} can be written as follows:
$E_{\rm c}\,\mbox{[MeV]} = 2.21\E^3 /R$ with $\E$ in GeV
and $R$ in millimeters.
For a 855 MeV projectile and for $R\gtrsim 10$ mm one
estimates $E_{\rm c} \lesssim 10^{-1}$ MeV.
Figure \ref{Figure09.fig} presents a low-energy part,
$E=20-500$ keV, of spectral distribution of radiation
emitted in the silicon crystals with different curvature radii.
As mentioned above, the spectral distributions correspond to the
emission cone that accounts practically for all radiation emitted
by the projectiles.
This results, in particular, in the increase of the
synchrotron-like radiation intensity with the curvature.
Figure \ref{Figure09.fig} shows that in the
in the energy range $E_{\rm c} \leq 250$ keV the most intensive
radiation is formed in the crystal with the smallest curvature
radius, $R=13.9$ mm.

The radiation emission within this energy range has not been
measured in Ref. \cite{BandieraEtAl:EPJC_v81_284_2021}.
However, the phenomenon of the synchrotron-like radiation
emitted by ultra-relativistic projectiles in different
oriented bent crystals deserves experimental investigation
especially in view of the ongoing efforts of the research and
technological communities towards design and practical
realization of novel gamma-ray crystal-based light sources
\cite{CLS-book_2022,TECHNO-CLS}.

\subsubsection{Volume reflection alignment \label{Spectra:VR}}

\begin{figure}[h]
\centering
\includegraphics[clip,width=12cm]{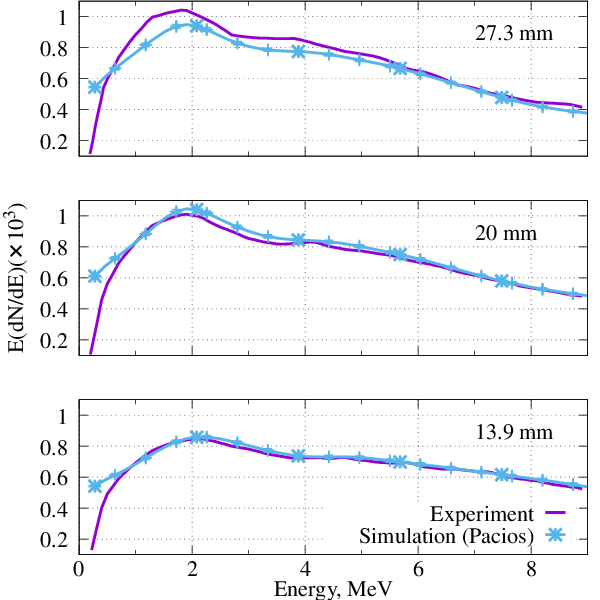}
\caption{
Spectral distribution of radiation emitted by 855 MeV electrons
passing through Si$(11\bar{1})$ crystal bent with the
curvature radius 27.3 mm (top), 20 mm (middle) and 13.9 mm
(bottom).
The incident beam geometry corresponds to the volume reflection
alignment with  $\theta=95$ mrad and
$\alpha=0.385$, 0.375 and 0.540 mrad in the top, middle and bottom
graphs, respectively.
Experimental data are from \cite{BandieraEtAl:EPJC_v81_284_2021}.
}
\label{Figure10.fig}
\end{figure}

\begin{figure}[h]
\centering
\includegraphics[clip,width=12cm]{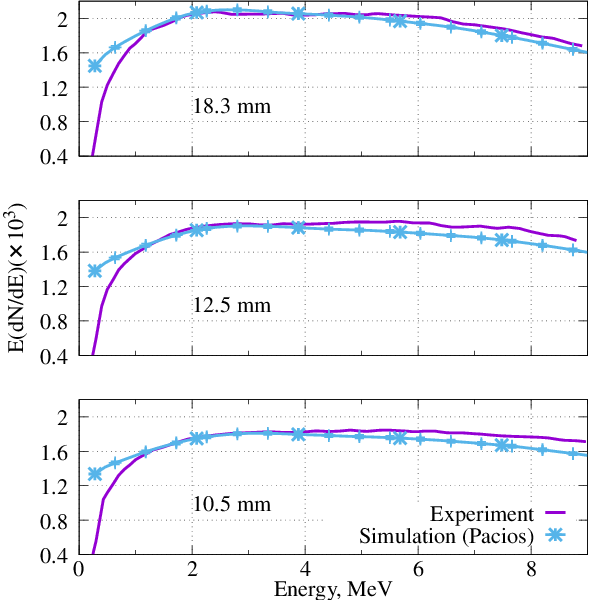}
\caption{
Same as in Fig. \ref{Figure10.fig} but for bent germanium
crystals.
The curvature radii are indicated in each graph.
The simulations were performed for the volume reflection
alignment with
$\alpha=0.410$ (top), 0.600 (middle) and 0.715 (bottom) mrad.
}
\label{Figure11.fig}
\end{figure}

Spectral distributions of radiation emitted by the electron beam
incident on the crystals at the volume reflection
alignment are presented in
Fig. \ref{Figure10.fig} and \ref{Figure11.fig}.
The results of simulations follow the experimentally measured
dependencies and reproduce the modifications in the spectra
observed in Ref. \cite{BandieraEtAl:EPJC_v81_284_2021}.

For both crystals the radiation intensity in the volume
reflection geometry
is smaller than under the channeling condition,
Figs. \ref{Figure06.fig} and \ref{Figure07.fig},
although the reduction is not dramatic so that in a broad range of
photon energies the intensities are noticeably higher that in the
amorphous targets.
The decrease in the $E(\d N/\d E)$ values is more pronounced for
larger curvature radius since the change in the geometry affects
the channeling efficiency which is higher in the crystals bent
with less curvature.
For the germanium crystal the decrease is quite moderate, varying
from approximately 20\% for $R=18.3$ mm down to 10\% for
$R=10.5$ mm.
For the silicon crystal the decrease is larger.
Comparing the peak values of the channeling radiation at
$E\approx 2$ MeV one observes the drop in the intensity by,
approximately, 40\% and 25\% for the largest and the smallest
curvature radius, respectively.

\section{Conclusions \label{Conclusions}}

By means of relativistic all-atom molecular dynamics implemented
in the \textsc{MBN Explorer} software package, the passage
of ultra-relativistic electrons through 15 $\mu$m thick silicon
and germanium bent crystals has been simulated.
The bending curvatures and the alignments of the 855 MeV electron
beam with respect to the crystallographic directions utilized in
the simulations correspond to those used in the experiments
\cite{SytovEtAl:EPJC_v77_901_2017,
BandieraEtAl:EPJC_v81_284_2021,%
HaurylavetsEtAl:EPJPlus_v137_34_2022}.
carried out at the MAMI facility.
The alignments considered include
(i) the channelling alignment, when the beam is aligned with the
tangent to the $(11\bar{1})$ planar direction at the
entrance, and (ii) the volume reflection regime.
For each case study considered, a sufficiently large number
of statistically independent trajectories has been generated.
Subsequent analysis of the trajectories resulted in the
calculation of the angular distribution of the electrons deflected
by a crystalline target, and the spectral distribution of
radiation emitted by the beam particles.
The simulations have been performed using two approximations for
the electron--atom interaction, due to Moli\'{e}re and to Pacios.
We have established that both approximations provide similar
results for the silicon crystal but in the case of the heavier
germanium crystal the Pacios approximation is more adequate.

Our results exhibit a high degree of agreement with the
experimental data especially in terms of the profiles of the
angular and spectral distributions as well as in the positions of
the maxima that are due to the channeling motion.
In case (i) main discrepancies have been revealed
in the angular distribution of electrons in the vicinity of the
over-barrier peak.
In case (ii) there are differences in the angular distributions'
maxima values.
The differences are more pronounced for the silicon crystal:
the experimental over-barrier maximum is 20\% higher than the
result of simulation whereas the channeling maximum is lower by
approximately the same amount.
For the Ge crystal the agreement is better: the maxima differ
by 10\%.

Two features, which were present in the experimental setup but
have not been accounted for in the present work, could be the
cause of the discrepancies.
First, in the simulations we have disregarded
the beam size of 105 $\mu$m and angular divergence of 21 $\mu$rad
divergence in the plane of the crystal bending
as quoted in Refs.
\cite{SytovEtAl:EPJC_v77_901_2017,BandieraEtAl:EPJC_v81_284_2021}.
Secondly,  when simulating the structure of quasi-mosaically
bent crystal, the secondary anticlastic curvature has not been
taken into account.
However, it has been demonstrated recently
\cite{SushkoEtAl:EPJD_v76_p236_2022} that the
accurate quantitative description of experimental results
on the beam deflection by such crystals can only be achieved if
one accounts for beam size and divergence as well as deduces
the entrance coordinate of the beam center in the plane of the
anticlastic bending.
In the cited paper the effect of the secondary anticlastic
curvature on the angular distribution of deflected electrons has
been analysed in connection with the experiment at the SLAC
facility \cite{WienandsEtAl:PRL_v114_074801_2015}.
We are planning to carry out similar analysis for the experiments
with the electron beam at MAMI.

\section*{Acknowledgements}

We acknowledge support by the European Commission
through the N-LIGHT Project within
the H2020-MSCA-RISE-2019 call (GA 872196)
and the EIC Pathfinder Project TECHNO-CLS
(Project No. 101046458).
V.V.H. has been partly supported by Grant BRFFI-MCTF,
No. F22MC-006.
We also acknowledge the Frankfurt Center for Scientific
Computing (CSC) for providing computer facilities.

\appendix
\section{Atomic and interplanar potentials \label{Appendix}}

For the sake of reference, we present numerical data
for atomic and (111) interplanar electron
potentials calculated
using the approximations due to Moli\'{e}re
\cite{Moliere}, Doyle and Turner  \cite{DoyleTurner1968},
and Pacios \cite{Pacios}.
Explicit formulae used in the calculations can be found in
Ref. \cite{KorolSushkoSolovyov:EPJD_v75_p107_2021}.
For brevity, below the abbreviations M, D-T and P are used
when referring to the approximations.

\begin{figure} [h]
\centering
\includegraphics[scale=0.5,clip]{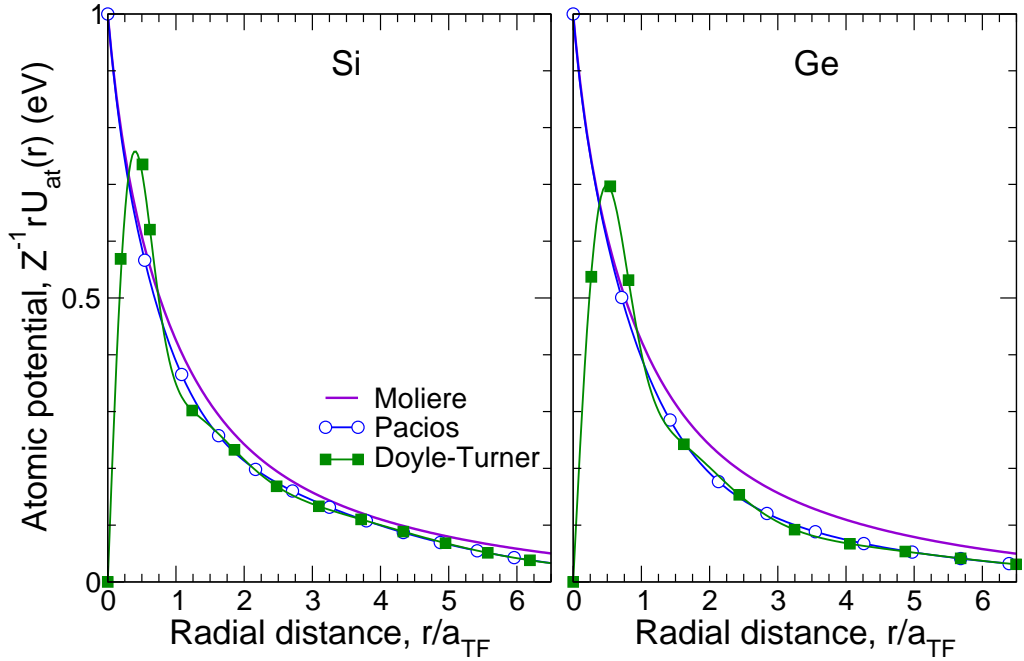}%
\caption{
Scaled atomic potentials $rU_{\rm a}(r)/Z$ versus
scaled radial distance $r/\aTF$
 calculated for Si (Z=14, $\aTF=0.194$ \AA) and
 Ge (Z=32, $\aTF=0.148$ \AA) atoms within
 the Moli\'{e}re, Pacios and Doyle-Turner approximations,
 as indicated in the common legend.
}
\label{Appendix:fig.01}
\end{figure}

Atomic potentials $U_{\rm a}$ of Si and Ge atoms are compared
in Figure \ref{Appendix:fig.01}.
The curves shown present the dependencies of the scaled
potential, $rU_{\rm a}(r)/Z$, on the radial distance measured
in the Thomas-Fermi radius $\aTF$, which is equal to 0.194 and
0.148 \AA{} for Si and Ge atoms, respectively.
In the domain $r\gtrsim 2\aTF$, the P and D-T
curves practically coincide lying below the
M curve.
This discrepancy is larger for the higher-$Z$ atom.
At small distances, the D-T parametrization
is incorrect whereas both M and P schemes provide the same
result that lead to the correct limit $rU_{\rm a}(r)/Z \to 1$
as $r\to 0$.

Within the continuous potential framework \cite{Lindhard}
the potential of an atomic plane in a crystal is obtained by
summing up potentials $U_{\rm at}$ of the atoms
in the plane assuming that they are distributed uniformly
within the plane.
This procedure also includes averaging of the atomic
positions due to thermal vibrations.
The inter-planar potential $U_{\rm pl}$ is calculated
as a sum of the potentials of individual planes.

\begin{figure} [h]
\centering
\includegraphics[scale=0.5,clip]{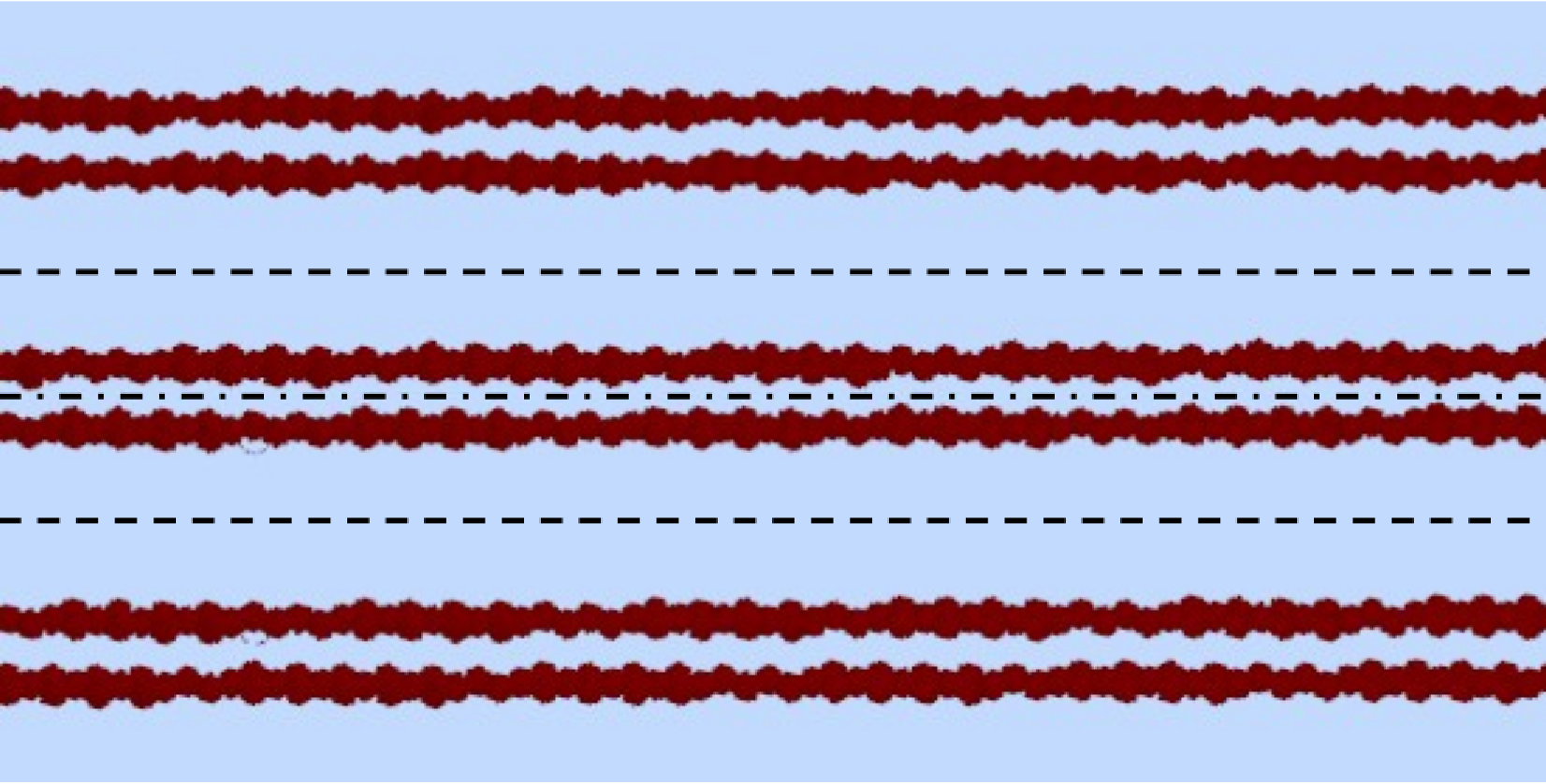}%
\caption{
Atoms (red circles) arranged in the (111) planes in a crystal
with the diamond cubic crystal structure.
Due to thermal vibrations the atoms are
shifted from the
nodal positions resulting in an uneven profile of the planes.
Two dashed lines mark the boundaries of the (111) channel
for a negatively charged projectile particle.
Dot-dashed line marks the channel's center-line.
The figure was rendered using the
multitask software toolkit
\textsc{MBNStudio} \cite{MBN_Studio_2019}.
}
\label{Appendix:fig.02}
\end{figure}

Silicon and germanium have the diamond cubic crystal structure
with lattice constants $a$ equal to 5.43 and 5.66 \AA,
respectively.
Distance between the (111) planes changes periodically from
the large distance equal to $3a/4$ to the small one $a/4$,
see illustrative Fig. \ref{Appendix:fig.02}.
For a negatively charged projectile (an electron, in particular)
the (111) channel includes two planes separated by $a/4$.
As a result, the continuous inter-planar potential as a function
of the distance $\rho$ from the channel's center
(dot-dashed line in the figure) has two symmetric minima
located at $\rho = \pm a/4$.

\begin{figure} [h]
\centering
\includegraphics[scale=0.5,clip]{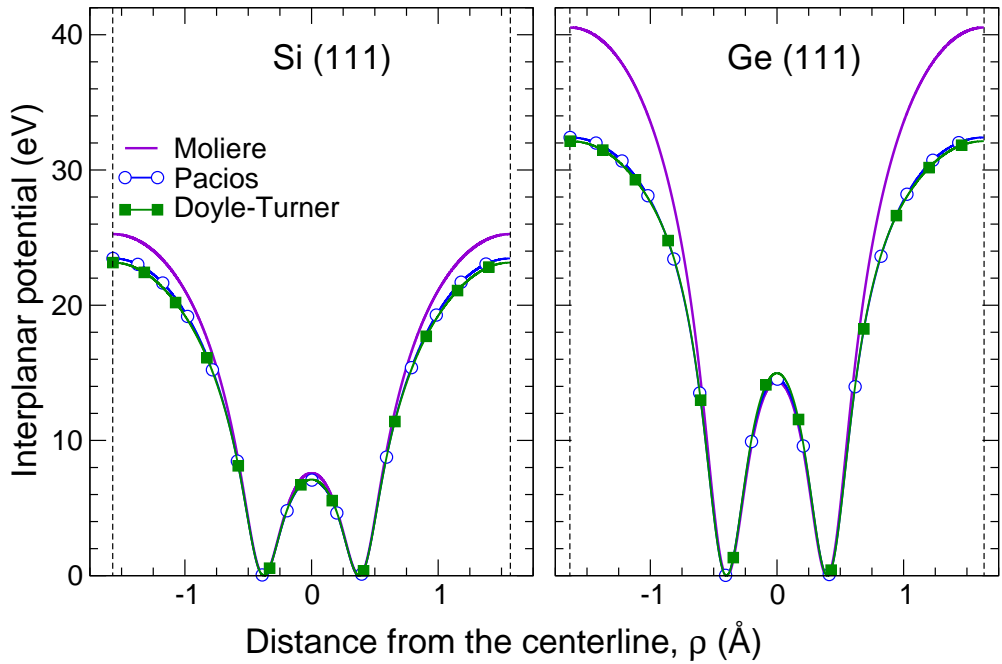}
\caption{
Electron (111) inter-planar potentials $U_{\rm pl}(\rho)$
in silicon (left) and germanium (right) crystals
calculated at $T=300\, ^{\circ}\mbox{C}$ using
the Moli\'{e}re, Pacios and Doyle-Turner approximations.
}
\label{Appendix:fig.03}
\end{figure}

Two graphs in Fig. \ref{Appendix:fig.03} compare the
M, P and D–T electron's inter-planar
potentials $U_{\rm pl}(\rho)$ in silicon and germanium crystals.
The uniform distribution of the atoms within a plane removes
the aforementioned incorrect feature of the Doyle-Turner
\textit{atomic} potential.
It is seen that in the planar case the P and D-T
approximations converge to practically the same results while
the M approximation produces higher values (10 \% and 25 \% for
silicon and germanium, respectively) for the depth of the
potential well.

\section*{References}



\begin{thebibliography}{99}

\bibitem{Lindhard}
       J. Lindhard,
       Influence of crystal lattice on motion of energetic
       charged particles.
       K. Dan. Vidensk. Selsk. Mat. Fys. Medd. \textbf{34}, 1 (1965).

\bibitem{BiryukovChesnokovKotovBook}
         V. M. Biryukov, Yu. A. Chesnokov, and V. I. Kotov,
         \textit{Crystal Channeling and Its Application at
         High-Energy Accelerators.}
         Springer Science \& Business Media (2013).
\bibitem{UggerhojRPM}
         U. I. Uggerh\o{}j,
         {The interaction of relativistic particles with strong
         crystalline fields.}
         \textit{Rev. Mod. Phys.} \textbf{77}, 1131 (2005).

\bibitem{MazzolariEtAl:PRL_v112_135503_2014}
         A. Mazzolari, E. Bagli, L. Bandiera, V. Guidi, H. Backe,
         W. Lauth, V. Tikhomirov, A. Berra, D. Lietti, M. Prest,
         E. Vallazza, and  D. De Salvador,
         Steering of a sub-{GeV} electron beam through planar
         channeling enhanced by rechanneling.
        \textit{Phys. Rev. Lett.} \textbf{112}, 135503 (2014).

\bibitem{MazzolariEtAl:EPJC_v78_720_2018}
    A. Mazzolari, M. Romagnoni, R. Camattari, E. Bagli, L. Bandiera,
    G. Germogli, ,  V. Guidi, G. Cavoto,
    Bent crystals for efficient beam steering of multi TeV-particle
    beams.
    {\textmd Si} crystal.
    \textit{Eur. Phys. J. C} \textbf{78}, 720 (2018).

\bibitem{WienandsEtAl:PRL_v114_074801_2015}
         U. Wienands, T.~W. Markiewicz, J. Nelson, R. J. Noble,
         J. L. Turner, U.~I. Uggerh{\o}j,  T.~N. Wistisen,
         E. Bagli, L. Bandiera, G. Germogli, V. Guidi,
         A. Mazzolari, R. Holtzapple, and M. Miller,
          Observation of Deflection of a Beam of Multi-GeV
          Electrons by a Thin Crystal.
          \textit{Phys. Rev. Lett.} \textbf{114}, 074801 (2015).

\bibitem{Scandale_EtAl-PRB_v692_p78_2010}
         W. Scandale, G. Arduini, R. Assmann, C. Bracco,
         S. Gilardoni, V. Ippolito, E. Laface, R. Losito,
         A. Masi, E. Metral, V. Previtali, S. Redaelli,
         M. Silari, L. Tlustos, E. Bagli, S. Baricordi,
         P. Dalpiaz, V. Guidi, A. Mazzolari et al.,
         First results on the SPS beam collimation with bent
         crystals.
         \textit{Phys. Lett.} \textbf{692B}, 78-82 (2010).

\bibitem{Scandale_EtAl-NIMB_v446_p15_2019}
         W. Scandale, G. Arduini, F. Cerutti, M. Garattini, S.
         Gilardoni, A. Lechner, R. Losito, A. Masi, D. Mirarchi,
         et al.,
         Focusing of 180 GeV/c pions from a point-like source
         into a parallel beam by a bent silicon crystal.
        \textit{Nucl. Instrum. Method  B} \textbf{446}, 15-18 (2019)

\bibitem{CLS-book_2022}
         A. V. Korol and A. V. Solov'yov.
         \textit{Novel Light Sources beyond Free Electron Lasers},
         Springer Nature Switzerland AG (2022).

\bibitem{SushkoEtAl:EPJ_v76_166_2022}
         G. B. Sushko, A. V. Korol and A. V. Solov'yov,
         {Extremely brilliant crystal-based light sources},
          \textit{Europ. Phys. J. D} \textbf{76}, 166 (2022).

\bibitem{AVK-AVS:NIMB_v537_p1_2023}
         A. V. Korol and A. V. Solov'yov,
         Atomistic modeling and characterizaion of light sources
         based on small-amplitude short-period periodically bent
         crystals.
          \textit{Nucl. Instrum. Methods B} \textbf{537}, 1 (2023).

\bibitem{Kumakhov:PL_v57A_17_1976} 
        M. A. Kumakhov.
        On the theory of electromagnetic radiation of charged particles
        in a crystal.
        Phys. Lett. \textbf{57A}, 17-18 (1976).

\bibitem{KaplinVorobiev:PLA_v67_p135_1978}
         V. V. Kaplin and S. A. Vorobiev,
         {On the electromagnetic radiation of channeled particles
         in a curved crystal.}
         \textit{Phys. Lett.} \textbf{67A}, 135 (1978).

\bibitem{TaratinVorobiev:NIMB_v31_p551_1988}
         A. M. Taratin and S. A. Vorobiev,
         {Radiation of high-energy positrons channeled in bent
         crystals.}
         \textit{Nucl. Instrum. Meth. B} \textbf{31}, 551 (1988).

\bibitem{KSG1998}
         Korol, A. V., Solov'yov, A. V. and Greiner, W.
         Coherent radiation of an ultrarelativistic charged particle
         channeled in a periodically bent crystal.
         \textit{J. Phys. G: Nucl. Part. Phys.} \textbf{24}, L45
         (1998).

\bibitem{SecondEdition}
        Andrey V. Korol, Andrey V. Solov’yov, and Walter Greiner.
       \textit{Channeling and Radiation in Periodically Bent
        Crystals},
        Second Edition.
        {Springer Series on Atomic, Optical, and Plasma Physics}
        \textbf{69},
         Springer-Verlag, Berlin Heidelberg (2014).

\bibitem{KorolSushkoSolovyov:EPJD_v75_p107_2021}
         A. V. Korol, G. B. Sushko, A. V. Solov’yov,
         All-atom relativistic molecular dynamics simulations of
         channeling and radiation processes in oriented crystals.
         \textit{Europ. Phys. J. D} \textbf{75}, 107 (2021)

\bibitem{BackeLauth:NIMB_v335_p24_2015} %
        H. Backe and W. Lauth,
        Channeling experiments with sub-GeV electrons in flat
        silicon single crystals.
        \textit{Nucl. Instrum. Meth. B} \textbf{335}, 24 (2015).

\bibitem{Scandale_EtAl-EPJC_v79_99_2019}
         W. Scandale, L. S. Esposito, M. Garattini, R. Rossi,
         V. Zhovkovska, A. Natochii, F. Addesa, F. Iacoangeli,
         F. Galluccio, F. Murtas, A. G. Afonin, Yu. A. Chesnokov,
         A. A. Durum, V. A. Maisheev, Yu. E. Sandomirskiy,
         A. A. Yanovich, G. I. Smirnov, Yu. A. Gavrikov,
         Yu. M. Ivanov, M. A. Koznov, M. V. Malkov,
         L. G. Malyarenko, I. G. Mamunct, J. Borg, T. James,
         G. Hall, M. Pesaresi,
         Reduction of multiple scattering of high-energy
         positively charged particles during channeling in
         single crystals.
         \textit{Eur. Phys. J. C} \textbf{79}, 99 (2019)


\bibitem{BandieraEtAl:PRL_v115_025504_2015}
        L. Bandiera, E. Bagli, G. Germogli, V. Guidi,
        A. Mazzolari, H. Backe, W. Lauth,  A. Berra, D. Lietti,
        M. Prest, D. De Salvador, E. Vallazza, and V. Tikhomirov,
        Investigation of the electromagnetic radiation emitted by
        sub-GeV electrons in a bent crystal.
        \textit{Phys. Rev. Lett.} \textbf{115}, 025504 (2015)

\bibitem{WistisenEtAl:PR-AB_v19_071001_2016}%
        T. N. Wistisen, U. I. Uggerh\o{}j, U. Wienands,
        T. W. Markiewicz, R. J. Noble, B. C. Benson, T. Smith,
        E.  Bagli, L. Bandiera, G. Germogli, V. Guidi,
        A. Mazzolari, R. L. Holtzapple, and S. Tucker,
        Channeling, volume reflection, and volume capture
        study of electrons in a bent silicon crystal.
        \textit{Phys. Rev. Accel. Beams} \textbf{19}, 071001 (2016).

\bibitem{Bagli_EtAl-EPJC_v77_71_2017}
         E. Bagli, V. Guidi, A. Mazzolari, L. Bandiera,
         G. Germogli, A. I. Sytov, D. De Salvador, A. Berra,
         M. Prest, E. Vallazza,
         Experimental evidence of independence of nuclear
         de-channeling length on the particle charge sign.
         \textit{Eur. Phys. J. C} \textbf{77}, 71 (2017)

\bibitem{SytovEtAl:EPJC_v77_901_2017}
         A. I. Sytov, L. Bandiera, D. De Salvador,
         A. Mazzolari, E. Bagli, A. Berra, S. Carturan,
         C. Durighello, G. Germogli, V. Guidi, P. Klag, W. Lauth,
         G. Maggioni, M. Prest, M. Romagnoni, V. V. Tikhomirov,
         and E. Vallazza,
         Steering of Sub-GeV electrons by ultrashort Si and Ge
         bent crystals.
         \textit{Eur. Phys. J. C} \textbf{77}, 901 (2017).

\bibitem{Wienands_EtAl-NIMB_v402_p11_2017}
        U. Wienands, S. Gessner, M. J. Hogan, T. W. Markiewicz,
        T. Smith, J. Sheppard, U. I. Uggerh\o{}j, J. L. Hansen,
        T. N. Wistisen, E. Bagli, L. Bandiera, G. Germogli,
        A. Mazzolari, V. Guidi, A. Sytov, R. L. Holtzapple,
        K. McArdle, S. Tucker, B. Benson,
        Channeling and radiation experiments at SLAC.
        \textit{Nucl. Instrum Meth. B} \textbf{402}, 11 (2017)

\bibitem{BandieraEtAl:EPJC_v81_284_2021}
          L. Bandiera, A. Sytov, D. De Salvador,
          A. Mazzolari, E. Bagli, R. Camattari, S. Carturan,
          C. Durighello, G. Germogli, V. Guidi, P. Klag, W. Lauth,
          G. Maggioni, V. Mascagna, M. Prest, M. Romagnoni,
          M. Soldani, V. V. Tikhomirov, and E. Vallazza,
          Investigation on radiation generated by sub-GeV electrons in
          ultrashort silicon and germanium bent crystals.
          \textit{Eur. Phys. J. C} \textbf{81}, 284 (2021).

\bibitem{BackeEtAl:JPConfSer_v438_012017_2013}
         H. Backe, D. Krambrich, W. Lauth,
         K.~K. Andersen, J.~L. Hansen, U.~I. Uggerh\o{}j,
         Channeling and Radiation of Electrons in Silicon
          Single Crystals and Si$_{1-x}$Ge$_x$ Crystalline
          Undulators.
         \textit{J. Phys.: Conf. Ser.} \textbf{438}, 012017 (2013).

\bibitem{BagliEtAl:EPJC_v74_3114_2014}
         E. Bagli, L. Bandiera, V. Bellucci, A. Berra,
         R. Camattari, D. De Salvador, G. Germogli, V. Guidi,
         L. Lanzoni, D. Lietti, A. Mazzolari, M. Prest,
         V. V. Tikhomirov, and E. Vallazza,
         Experimental evidence of planar channeling in a
         periodically bent crystal,
         \textit{Eur. Phys. J. C} \textbf{74}, 3114 (2014).

 \bibitem{WistisenEtAl:PRL_v112_254801_2014}
         T. N. Wistisen,  K. K. Andersen, S. Yilmaz, R. Mikkelsen,
         J. L. Hansen, U. I. Uggerh\o{}j, W. Lauth, and H. Backe.
         Experimental realization of a new type of crystalline
         undulator,
         \textit{Phys. Rev. Lett.} \textbf{112}, 254801 (2014).

\bibitem{UggerhojWistisen:NIMB_v355_p35_2015}
        U.~I. Uggerh\o{}j, T.~N. Wistisen:
        Intense and energetic radiation from crystalline undulators,
       \textit{Nucl. Instrum. Meth. B} \textbf{355}, 35 (2015).

\bibitem{MBN_Explorer_2012}
         I.~A. Solov'yov, A.~V. Yakubovich, P.~V. Nikolaev,
         I. Volkovets, and A.~V. Solov'yov,
         {MesoBioNano Explorer -- A universal program for
         multiscale computer simulations of complex molecular
         structure and dynamics.}
         \textit{J. Comp. Phys.} \textbf{33}, 2412 (2012).

\bibitem{MBNExplorer_Book}
        I. A. Solov'yov,  A. V. Korol, and A. V. Solov'yov,
        \textit{Multiscale Modeling of Complex Molecular
        Structure and Dynamics with MBN Explorer.}
        Springer International Publishing, Cham, Switzerland (2017).

\bibitem{mbn-explorer-software}
     MBN Explorer and MBN Studio Software at
    http://mbnresearch.com/software-0

\bibitem{MBN_Studio_2019}
         G. B. Sushko, I. A. Solov'yov, and A. V. Solov'yov,
        {Modeling MesoBioNano systems with MBN Studio made
        easy}.
        \textit{J. Mol. Graph. Model.} \textbf{88}, 247 (2019).

\bibitem{SalvadorEtAl:JINST_v13_C04006_2018} 
        D. De Salvador, S. Carturan, A. Mazzolari, E. Bagli,
        L. Bandiera, C. Durighello, G. Germogli, V. Guidi,
        P. Klag, W. Lauth, G. Maggioni, M. Romagnoni, and A. Sytov,
        Innovative remotely-controlled bending device for thin silicon and germanium crystals.
        \textit{JINST}, \textbf{13}, C04006 (2018).

\bibitem{HaurylavetsEtAl:EPJPlus_v137_34_2022}
        V.~V. Haurylavets, A. Leukovich, A. Sytov, L. Bandiera,
        A. Mazzolari, M. Romagnoni, V. Guidi, G.~B. Sushko,
        A.~V. Korol, and A.~V. Solov’yov,
        MBN Explorer atomistic simulations of 855 MeV electron
        propagation and radiation emission in oriented silicon bent crystal:
        theory versus experiment,
        \textit{Eur. Phys. J. Plus} \textbf{137}, 34 (2022).

\bibitem{MBN_ChannelingPaper_2013}
         G.~B. Sushko, V.~G. Bezchastnov, I.A. Solov'yov, A.~V. Korol,
         W. Greiner, and A.~V. Solov'yov,
         Simulation of ultra-relativistic electrons and positrons
         channeling in crystals with MBN Explorer.
         \textit{J. Comp. Phys.} \textbf{252}, 404 (2013).


\bibitem{Moliere}
         G. Moli\`ere,
         Theorie der Streuung schneller geladener
         Teilchen I: Einzelstreuung am abgeschirmten Coulomb-Feld.
        \textit{Z. f. Naturforsch} \textbf{A2}, 133 (1947).

\bibitem{Pacios}
         L. Fernandes Pacios,
         Analytical Density-Dependent Representation of
         Hartree-Fock Atomic Potentials.
         \textit{J. Comp. Chem.} \textbf{14}, 410 (1993).

\bibitem{DoyleTurner1968}
    P. A. Doyle and P. S. Turner,
    {Relativistic Hartree-Fock X-ray and Electron Scattering
    Factors.}
    \textit{Acta Cryst. A} \textbf{24}, 390 (1968).


\bibitem{SushkoEtAl:JPConfSer_v438_012019_2013}
         G.~B. Sushko, V.~G. Bezchastnov, A.~V. Korol,
         Walter Greiner, A.~V. Solov'yov,  R.~G. Polozkov, and
         V.~K. Ivanov,
         Simulations of electron channeling in bent silicon
         crystal.
         \textit{J. Phys. Conf. Ser.} \textbf{438}, 012019 (2013).

\bibitem{SushkoEtAl:JPConfSer_v438_012018_2013}%
         G.~B. Sushko, A.~V. Korol, Walter Greiner, and
         A.~V. Solov'yov,
         Sub-GeV Electron and Positron Channeling in Straight,
         Bent and Periodically Bent Silicon Crystals.
         \textit{J. Phys. Conf. Ser.} \textbf{438}, 012018 (2013).

\bibitem{PolozkovEtAl:EPJD_v68_268_2014}
         R.~G. Polozkov, V.~K. Ivanov, G.~B. Sushko, A.~V. Korol,
         and A.~V. Solov'yov,
         Radiation emission by electrons channeling in bent
         silicon crystals.
         \textit{Eur. Phys. J. D} \textbf{68}, 268 (2014).

\bibitem{Sushko:Thesis_2015}
         G. B. Sushko,
         \textit{Atomistic Molecular Dynamics Approach for
         Channeling of Charged Particles in Oriented Crystals}
         (Doctoral dissertation),
         Goethe-Universit\"{a}t, Frankfurt am Main (2015).

\bibitem{IvanovEtAl:JETPLett_v81_p977_2005}
         Y. Ivanov, A. Petrunin, and V. Skorobogatov,
         {Observation of the elastic quasi-mosaicity effect
         in bent silicon single crystals.}
         \textit{JETP Lett.} \textbf{81}, 977 (2005).

\bibitem{GuidiEtAl:JPD_v42_182005_2009}
         V. Guidi, A. Mazzolari, D. De Salvador, and A. Carnera,
         Silicon crystal for channelling of negatively charged particles.
         \textit{J. Phys. D: Appl. Phys.} \textbf{42}, 182005 (2009).

\bibitem{CamattariEtAl:JAC_v489_p977_2015}
         R. Camattari, V. Guidi, V. Bellucci, and A. Mazzolari,
         The quasi-mosaic effect in crystals and its application
         in modern physics,
         \textit{J. Appl. Cryst.} \textbf{48}, 977 (2015).

\bibitem{BackeEtAl:NIMB_v266_p3835_2008}
         H. Backe, P. Kunz, W. Lauth, and A. Rueda,
         Planar channeling experiments with electrons at the 855
         MeV Mainz Microtron MAMI.
         \textit{Nucl. Instrum. Meth. B} \textbf{266}, 3835
         (2008).

\bibitem{SytovTikhomirov_NIMB_v355_p383_2015}
        A. Sytov and V. Tikhomirov,
        CRYSTAL simulation code and modeling of coherent effects
        in a bent crystal at the LHC.
       \textit{Nucl. Instrum. Methods B} \textbf{355}, 383 (2015).

\bibitem{TaratinVorobiev:PLA_v115_p398_1986}
         A. M. Taratin and S. A. Vorobiev,
         {Volume trapping of protons in the channeling regime in a bent
         crystal.}
         \textit{Phys. Lett.} \textbf{115A}, 398 (1986).

\bibitem{TaratinVorobiev:PLA_v119_p425_1987}
         A. M. Taratin and S. A. Vorobiev,
         {Volume reflection of high-energy charged particles in quasi-channeling states in bent crystals.}
         \textit{Phys. Lett.} \textbf{119A}, 425 (1987).

\bibitem{ShenEtAl:NIMB_v424_p26_2018}
         H. Shen, Q. Zhao, F.~S. Zhang, G.~B. Sushko,
         A.~V. Korol, and A.~V. Solov'yov,
         Channeling and radiation of 855 MeV electrons and
         positrons in straight and bent tungsten (110) crystals.
          \textit{Nucl. Instrum. Meth. B} \textbf{424}, 26 (2018).

\bibitem{Jackson}
         J.D. Jackson,
         {\it Classical Electrodynamics}
         (Wiley, Hoboken, 1999)

\bibitem{TECHNO-CLS}
         http://www.mbnresearch.com/TECHNO-CLS/Main.

\bibitem{SushkoEtAl:EPJD_v76_p236_2022}
		G.~B. Sushko,  A.~V. Korol, A.~V. Solov'yov,
        Ultra-relativistic electron beams deflection by quasi-
        mosaic crystals.
        \textit{Eur. Phys. J. D} \textbf{76}, 236 (2022).

\end{thebibliography}
\end{document}